# A study of Pt, Rh, Ni and Ir dispersion on anatase TiO$_2$(101) and the role of water

Lena Puntscher, Kevin Daninger, Michael Schmid, Ulrike Diebold and Gareth S. Parkinson*

Institute of Applied Physics, TU Wien, Vienna, Austria

*parkinson@iap.tuwien.ac.at

**Abstract**

Understanding how metal atoms are stabilized on metal oxide supports is important for predicting the stability of "single-atom" catalysts. In this study, we use scanning tunnelling microscopy (STM) and x-ray photoelectron spectroscopy (XPS) to investigate four catalytically active metals – Platinum, Rhodium, Nickel and Iridium – on the anatase TiO$_2$(101) surface. The metals were vapor deposited at room temperature in ultrahigh vacuum (UHV) conditions, and also with a background water pressure of $2×10^{-8}$ mbar. Pt and Ni exist as a mixture of adatoms and nanoparticles in UHV at low coverage, with the adatoms immobilized at defect sites. Water has no discernible effect on the Pt dispersion, but significantly increases the amount of Ni single atoms. Ir is highly dispersed, but sinters to nanoparticles in the water vapor background leading to the formation of large clusters at step edges. Rh forms clusters on the terrace of anatase TiO$_2$(101) irrespective of the environment. We conclude that introducing defect sites into metal oxide supports could be a strategy to aid the dispersion of single atoms on metal-oxide surfaces, and that the presence of water should be taken into account in the modelling of single-atom catalysts.

## 1. Introduction

Precious metal nanoparticles supported on metal oxides are used for chemical conversions in heterogeneous and electrocatalysis. Reducing the size of the particles increases the fraction of atoms present at the nanoparticles' surface, and thus the per-atom efficiency. When the particles enter the subnano regime, quantum size effects can affect the catalytic activity.[1, 2] In the ultimate limit, single metal atoms can be anchored directly onto the oxide support, and so-called single-atom catalysis (SAC) has emerged as a key strategy in heterogeneous and electrocatalysis in the last decade.[3-7]

Unravelling how catalytically active metal atoms bind to the metal oxide support and interact with reactants is essential for understanding their properties. The local coordination environment has been shown to strongly influence the reactivity and stability of SACs,[8-11] but the structural details of the active sites are difficult to obtain from experiment. This is partly due to the structural inhomogeneity of powder supports, and partly due to the limitations of analytical techniques. In the absence of this critical information, the reaction mechanism is typically modelled computationally assuming an idealized low-index facet of the support material with the catalyst atom located at a high-symmetry site. Such models almost certainly do not represent the active catalyst, particularly for electrochemical applications, because the presence of water and/or hydroxyl groups is neglected.

One approach to investigate the validity of the assumptions made in the computational modelling of SACs is to synthesize analogous systems experimentally. Such experimental modelling is achieved using single-crystalline metal-oxide supports where the atomic structure is well known. The metal of interest is evaporated directly onto the pristine surface in ultrahigh vacuum, which allows the most stable adsorption site to be determined. One can also selectively introduce molecules that might affect the stability of the system, and determine their individual impact unambiguously. For systems ultimately utilized in an aqueous solution, water is the obvious candidate. Recently, we demonstrated that Rh atoms sinter rapidly after deposition on a pristine α-Fe$_2$O$_3$($1\bar{1}02$) model support in ultrahigh vacuum (UHV) at room temperature but are stabilized as "single atoms" when the same experiment is performed with 10$^{-8}$ mbar water in the background pressure. The enhanced dispersion occurs because the adatoms are stabilized by additional coordination to two OH ligands.[12] On the other hand, adsorbates can also induce sintering, as observed for Pd/Fe$_3$O$_4$(001) in the presence of CO.[13]

In this work, we turn our attention to TiO$_2$ as a model support. The thermodynamically stable rutile phase of TiO$_2$, especially the (110) surface, has been widely investigated in surface science studies.[14, 15] In SAC, the anatase polymorph (a-TiO$_2$) is of particular interest because it becomes more stable in the nanoparticle form typically used as a support [16]. The reactivity of a-TiO$_2$-supported SAC systems has been heavily investigated in recent years [17-23]. DeRita et al. [24], for example, have convincingly demonstrated that Pt adatoms are active for CO oxidation. The possibility that agglomerates might be responsible for the observed activity was ruled out by using a very low Pt loading; each support particle hosted on average just one Pt atom. DFT-based calculations accompanying the experiments suggested that Pt atoms are probably not stable on the bare a-TiO$_2$(101) surface. Moreover, a single Pt adatom on top of the bare a-TiO$_2$(101) surface also failed to reproduce the experimentally observed binding energy and vibrational frequency of Pt-adsorbed CO molecules. These benchmark parameters were best reproduced when the Pt atoms were assumed to be coordinated to two additional oxygen atoms originating from hydroxyl groups on the surface.[24] Another study by the same group [25] reported that the coordination of Rh adatoms on a-TiO$_2$ is sensitive to the composition of the reducing gas that was used to activate the catalyst. When Rh was pre-treated in CO at 300 °C and further exposed to CO, Rh(CO)$_2$ species formed with Rh being bound to two O$_{2c}$ from the lattice. (For a sketch of the a-TiO$_2$ surface structure and the adsorption site, see Fig. 1, below) When the system was pre-treated with H$_2$ at 100 °C, hydroxyls formed. These coordinated to the Rh(CO)$_2$ species by adding an additional neighbouring surface OH group which substantially changed the CO binding energy. It was concluded from CO FTIR-TPD and DFT that the presence of hydroxyl groups can alter the local metal coordination and molecular desorption significantly.[25]

Here, we present a surface science investigation study of four different metals – Pt, Rh, Ni and Ir – vapor-deposited directly onto an a-TiO$_2$(101) single crystal support at room temperature in UHV. We find that Ir is the only metal that exhibits atomic dispersion under UHV conditions. However, the presence of water de-stabilizes the Ir adatoms, which leads to the formation of large clusters anchored at step edges. Pt, Ni, and Rh all form mostly clusters even at very low coverages, suggesting diffusion is facile on the regular terrace at room temperature. For Pt and Ni, small protrusions are observed in the STM images that we tentatively assign as isolated adatoms immobilized at defects.

## 2. Experimental Methods

Room-temperature scanning tunnelling microscopy (STM) was performed in a two-vessel UHV chamber consisting of a preparation chamber (base pressure p < 10$^{-10}$ mbar) and an analysis chamber (p < 5×10$^{-11}$ mbar). The analysis chamber is equipped with a nonmonochromatic Al Kα X-ray source (VG) and a SPECS Phoibos 100 analyzer for XPS, and an Omicron μ-STM. The STM measurements (positive sample bias, empty states) were conducted in constant current mode with an electrochemically etched W tip. The natural a-TiO$_2$(101) single crystal was prepared in UHV by sputtering (Ar$^+$, 1 keV, 10 min) and annealing (610 °C, 20 min). Every fifth cycle the sample was annealed at 500 °C for 20 min in 5×10$^{-7}$ mbar of O$_2$ and then in UHV at 610 °C.[26] Pt, Rh, Ni and Ir were deposited using an e-beam evaporator (FOCUS), with the flux calibrated using a water-cooled quartz microbalance (QCM). One monolayer (ML) is defined as one metal atom per surface unit cell. (The areal density of unit cells is 5.15 × 10$^{18}$ m$^{-1}$) The STM images were corrected for distortion and creep of the piezo scanner as described in ref [27]. The gray scale of each image is set individually to ensure that the possible adatoms and other small adsorbates are easily distinguishable. Furthermore, a background subtraction is done by setting the iron atoms of the support to an apparent height of zero.

## 3. Results
### 3.1 The as-prepared anatase TiO$_2$(101) surface

Figure 1a shows STM images of the a-TiO$_2$(101) surface after several cleaning cycles. As is typical for this well-studied surface, a clean sample exhibits rows of bright, oval-shaped protrusions running in the [010] direction. These are attributed to the surface Ti$_{5c}$ and O$_{2c}$ atoms shown in Figures 1c, d. [28] The [10-1] direction cannot be easily determined from these images; it was inferred from the preferred step directions.[28] Isolated dark features (highlighted with a blue arrow in Figure 1a) between the rows are inhomogeneously distributed over the surface. These have previously been attributed to extrinsic Nb dopants, which are often present in natural anatase TiO$_2$ samples.[29] Our XPS survey spectra did not show a peak that would allow us to confirm or debunk this assignment, likely because their average coverage is too low (0.02–0.03 ML as measured by STM). For what follows, it is important to note that surface oxygen vacancies (V$_O$S) are not present at the surface of a-TiO$_2$(101). Even when formed artificially, they quickly diffuse to the subsurface at room temperature [30]. This is in stark contrast to rutile TiO$_2$(110), where V$_O$ sites are prevalent and active sites for adsorption [14].

Figure 1b shows the surface after 2 hours of exposure to the residual gas of the preparation chamber at room temperature (with the evaporator turned on but the shutter closed). Bright protrusions are observed; there appear identical to those observed after water adsorption in low-temperature studies [31]. Since water is known to desorb from regular a-TiO$_2$(101) surface sites below 250 K [32], we conclude these water molecules are adsorbed at surface defects. The concentration agrees with that of the dark defects highlighted in Figure 1a. Interestingly, the water molecules are mobile at room temperature (Figure 2), but do not leave a visible defect behind when diffusing. This suggests that water and defect probably diffuse together, which makes it unlikely that the defect can be a cation substituting Ti in the anatase lattice. It could conceivably be an interstitial lattice species, or perhaps a surface site above a subsurface defect such as an oxygen vacancy. The images also exhibit a low concentration of molecular O$_2$ [33] species, which are in the residual gas as left-over from the oxidation step during sample preparation. This species is also most likely bound at defects, and a few examples are highlighted in orange and marked as (O$_2$)$_{extr}$ in Figure 1b, consistent with the labelling in ref. [29]. These (O$_2$)$_{extr}$ are also mobile at room temperature, and in a rare case we observed one of them to hop onto a dark defect, without leaving a similar defect behind (Figure 3). This suggests that there may be defect sites that are not visible in STM images, where adsorbates can bind more strongly than at regular surface sites. Overall, these data show that the regular anatase surface is inert at room temperature, but that defects (both visible and invisible in STM) can act as binding sites for molecular adsorbates. In what follows, we will show that these defects can also stabilize metal adatoms.

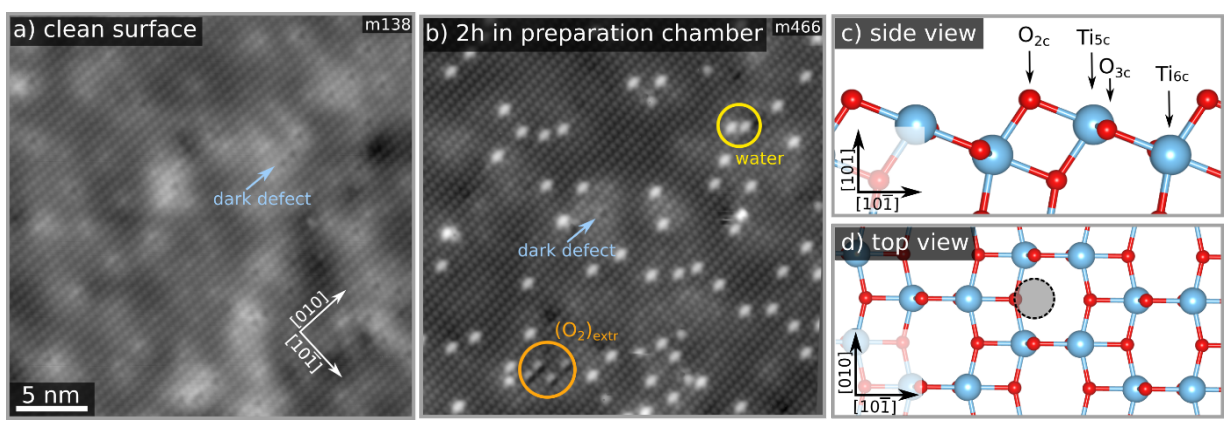

Figure 1: The as-prepared a-TiO$_2$(101) surface.  STM images (a) of the a-TiO$_2$(101) surface acquired shortly after preparation ($V_{sample}$ = +1 V, $I_{tunnel}$ = 0.3 nA), and (b) after keeping the sample in the preparation chamber for 2 hours ($V_{sample}$ = +1 V, $I_{tunnel}$ = 0.3 nA). In (a), dark defects (previously attributed to Nb dopants [29]) are highlighted in light blue. After exposing the surface to the residual gas in the preparation chamber for 2 hours (b) water molecules (highlighted in yellow) as well as O$_2$ molecules (highlighted in orange) adsorb at the surface, likely at defect sites [31]. Panels c) and d) show a model of the a-TiO$_2$(101) surface in which the O atoms are coloured red, and the Ti atoms are coloured blue. The most stable site for Pt and Rh adatom adsorption computed by prior studies is located between two surface O$_{2c}$ atoms (marked by the grey circle).[22, 34-36]

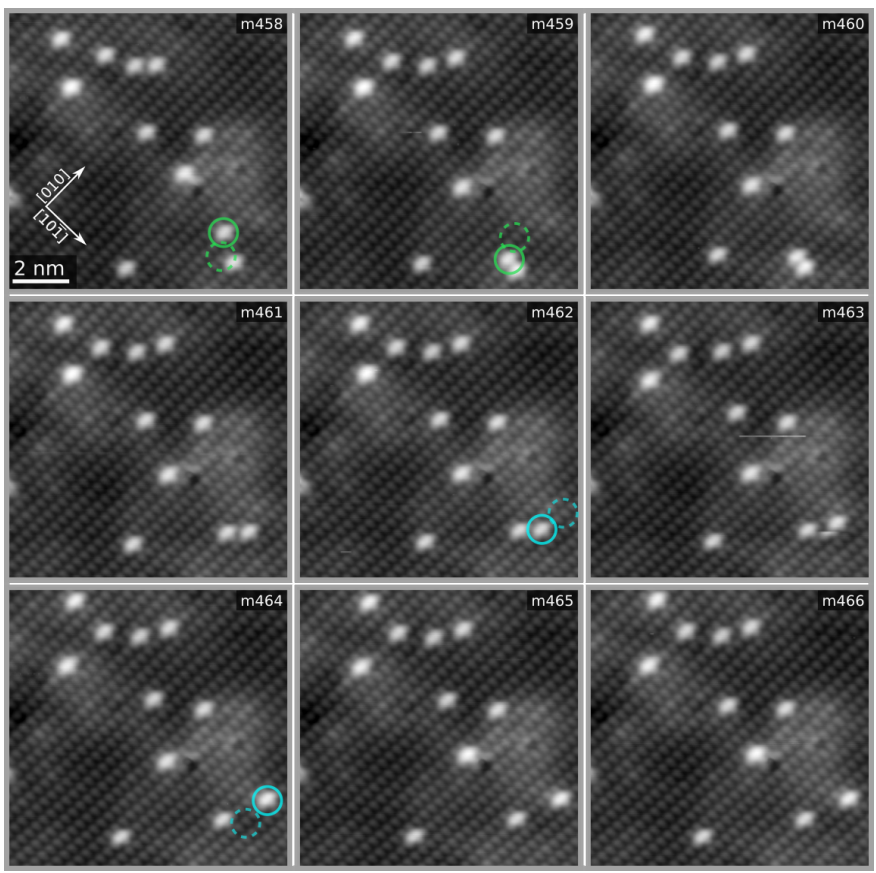

Figure 2: Diffusion of water on the a-TiO$_2$(101) surface. Water was adsorbed from the residual gas in the preparation chamber over the course of two hours at room temperature. Sequential STM images ($V_{sample}$ = +1 V, $I_{tunnel}$ = 0.3 nA, ≈ 300 nm/s) show that the water molecules can move along and across the [010] direction. Movements across the [010] direction are shown in green, movement along the [010] are shown in cyan. The full circles mark the current position of the water molecule, the dashed circle the position it has after respectively before the movement.

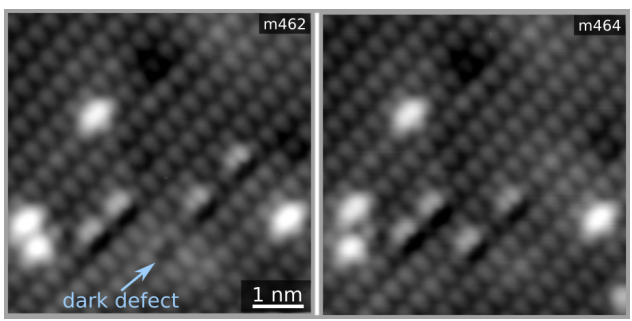

Figure 3: Sequential STM images acquired after keeping the sample in the preparation chamber for 2 hours at room temperature ($V_{sample}$ = +1 V, $I_{tunnel}$ = 0.3 nA). In the second image, an adsorbed $O_2$ molecule has moved to occupy a dark defect, leaving behind an apparently regular lattice site.

**3.2 Pt/anatase TiO$_2$(101)**

A previous STM study of the Pt/a-TiO$_2$(101) system [34] revealed that small clusters form predominantly on the terrace, with some species tentatively assigned to adatoms. Our data (Fig. 4a for a coverage of 0.05 ML) are similar to those in presented in ref. [34], and we also observe the coexistence of larger clusters and smaller features that have a uniform apparent height of 150-160 pm. At a lower coverage of 0.01 ML (Figure 4b), the density and size of the clusters is lower, and the 150-160 pm species are again observed. These smallest Pt species are easily distinguished from adsorbed water by their larger apparent height at our imaging conditions (60-80 pm for water, see figure 4c for a comparison), and because they are immobile in room-temperature STM movies. Given their relatively small apparent height, we tentatively assign these protrusions to Pt$_1$ species. From the experiment alone, we cannot discount that these species could be dimers (or trimers) if such species were significantly more stable. For the higher coverage, (0.05 ML) ≈7 % of the deposited Pt (according to the QCM calibration) can be attributed to possible single atoms, whereas at 0.01 ML this increases to ≈17 %.

Figure 5a shows a high-resolution image, in which orange dots mark the approximate position of surface Ti$_{5c}$ atoms. Assuming that the substrate maxima imaged by STM are closer to the Ti$_{5c}$ than the O$_{2c}$, the Pt-related protrusion is close to the position predicted by DFT calculations in ref. [34] (in between two adjacent O$_{2c}$ atoms, grey circles in Fig. 1d). We also note that the Pt adsorption site is equivalent to the sites where the dark defects are seen in STM (Figure 5f), so it is possible that these defects help stabilize the Pt atoms. Adsorbed water and O$_2$ are also labelled in Figs 4a and 4b for ease of comparison.

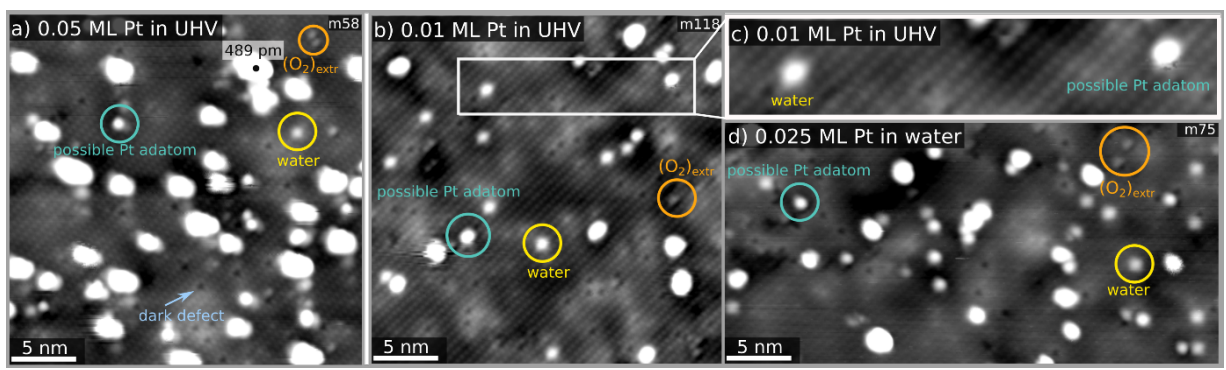

Figure 4: Pt on a-TiO$_2$(101). (a) After deposition of 0.05 ML Pt ($V_{sample}$ = +1.5 V, $I_{tunnel}$ = 0.3 nA), (b) 0.01 ML Pt ($V_{sample}$ = +1.5 V, $I_{tunnel}$ = 0.3 nA). (c) A magnified section with a water molecule and a possible Pt adatom. (d) After depositing 0.025 ML Pt in a background of 2×10$^{-8}$ mbar water vapor ($V_{sample}$ = +2 V, $I_{tunnel}$ = 0.3 nA). Possible Pt adatoms are highlighted in cyan. In (a) the cluster with the highest apparent height is marked. It measures 489 pm at its highest point.

Figure 4d shows an STM image of Pt deposited in a water vapor background of 2 × 10$^{-8}$ mbar. Again, a mixture of clusters and possible adatoms is observed, and the ratio of clusters and possible single atoms is comparable to that obtained in UHV. We thus conclude that water has no significant effect on the dispersion of Pt on the a-TiO$_2$(101) terraces, at least in this low-pressure regime.

Figure 5b shows that the possible Pt adatom adsorption site is the same, independent of whether deposition was done in a water vapor background or in UHV.

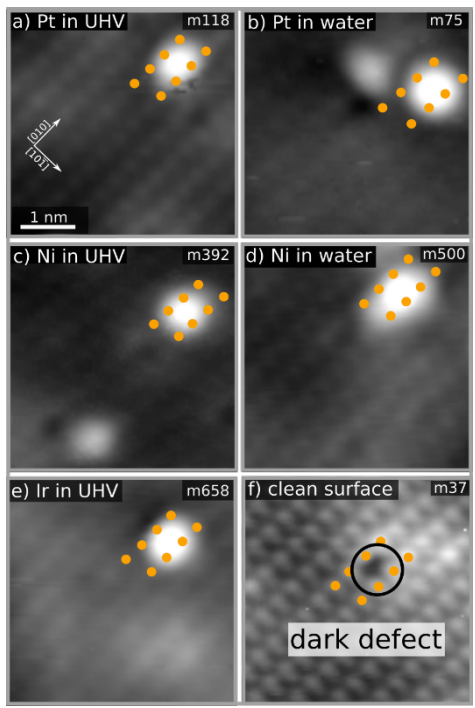

Figure 5: Enlarged STM images of the adatoms, to determine the extract adsorption site of all metal atoms deposited in UHV (a, c, e) and in a background of 2×10$^{-8}$ mbar water vapor (b, d) and of a dark defect (f) The orange dots mark the approximate positions of surface Ti$_{5c}$ atoms.

**3.3 Rh/anatase TiO$_2$(101)**

Figure 6 shows the a-TiO$_2$(101) surface after deposition of 0.02 ML Rh (a) in UHV and (b) in a water vapor background of 2×10$^{-8}$ mbar. Unlike Pt, Rh forms exclusively small clusters on the surface despite the presence of the dark defects. We did not observe any features that we would attribute to single atoms, irrespective of the environment. We conclude that Rh$_1$ species are not stable on the a-TiO$_2$(101) surface at room temperature under our conditions. This is similar to our experience with r-TiO$_2$(110) [37], where Rh$_1$ species were found to sinter already at 150 K.

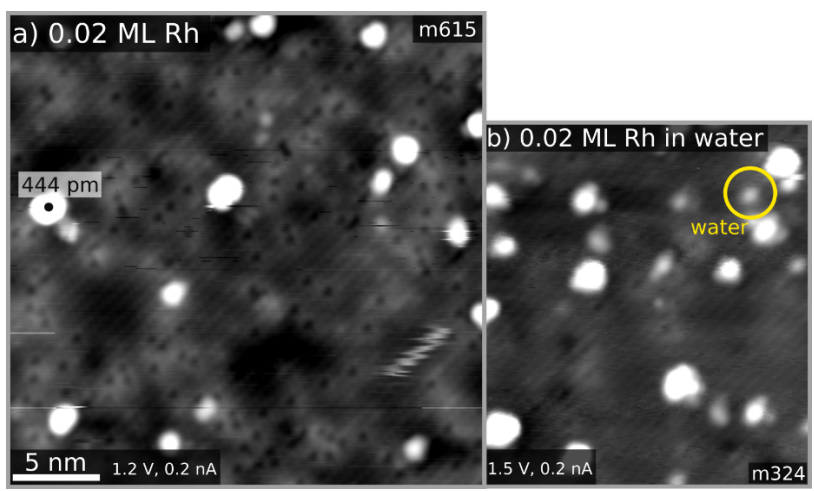

Figure 6: STM results of Rh on a-TiO$_2$(101). (a) After deposition of 0.02 ML Rh in UHV ($V_{sample}$ = +1.2 V, $I_{tunnel}$ = 0.2 nA) and (b) 0.02 ML Rh in a water vapor background of 2×10$^{-8}$ mbar ($V_{sample}$ = +1.5 V, $I_{tunnel}$ = 0.15 nA). In (a) the cluster with the highest apparent height is marked. It measures 444 pm at its highest point.

### 3.4 Ni/anatase TiO$_2$(101)

Figure 7 shows the surface after deposition of 0.02 ML Ni in a) UHV and a water vapor background of 2×10$^{-8}$ mbar. Like Pt, Ni forms a mixture of clusters and small, uniform features that could be attributed to adsorbed single atoms. The coverage of these small species is relatively high: assuming that they are Ni$_1$ they would account for ≈20 % of the deposited Ni, with the rest contained within larger clusters. The smallest species are easily distinguished from adsorbed water, partly by their apparent height (150-170 pm), and also because they are immobile on the a-TiO$_2$(101) surface at room temperature. Thus, in analogy to Pt, we presume that the smallest Ni species are most likely trapped at defect sites. Figures 5c and d show that these species are adsorbed at a different location on the surface than the features attributed to Pt atoms.

After deposition in a water background of 2×10$^{-8}$ mbar, the concentration of the Ni$_1$ species doubles from 20 % of the deposited Ni to 40 %. There is no discernible difference between the protrusions in the two experiments, so it seems that water has a significant effect on the dispersion of Ni and may play a role in stabilizing Ni at defect sites.

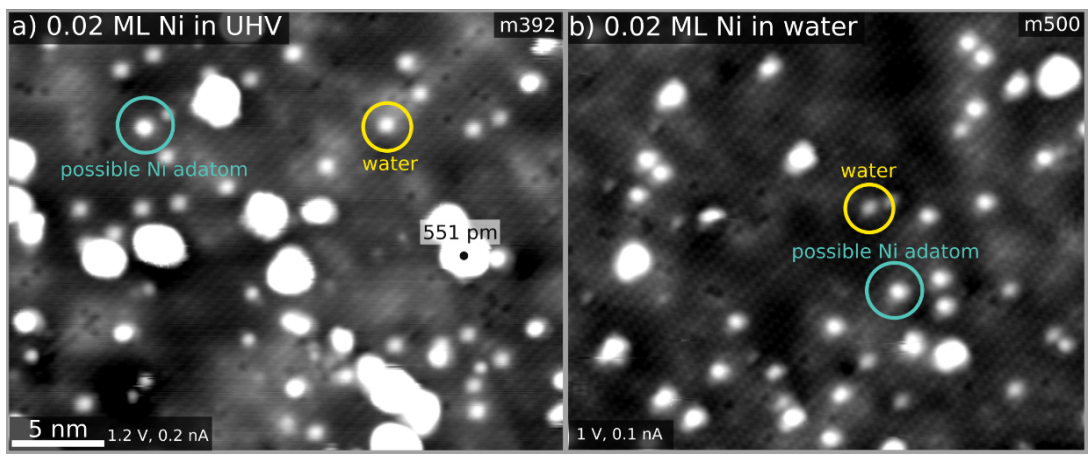

Figure 7: STM results of Ni on a-TiO$_2$(101). 0.02 ML Ni has been deposited (a) in UHV ($V_{sample}$ = +1.2 V, $I_{tunnel}$ = 0.15 nA) and (b) in a water vapor background of 2×10$^{-8}$ mbar ($V_{sample}$ = +1 V, $I_{tunnel}$ = 0.1 nA). The density of possible Ni adatoms doubles in the presence of water. In (a) the cluster with the highest apparent height is marked. It measures 551 pm at its highest point.

### 3.5 Ir/anatase TiO$_2$(101)

The last metal investigated in this study was Ir. Figure 8 shows 0.02 ML Ir deposited in UHV. Unlike Pt, Rh and Ni, Ir forms mostly uniform features with an apparent height of 130-160 pm. All these features occupy the same site on the surface and are immobile at room temperature. Like Pt, Ir is located approximately between two O$_{2c}$ surface atoms (Fig. 5). We assign these features to single Ir adatoms, which appear at a coverage of 0.011 ML on the surface. In addition to the single atoms, a small number of clusters can also be recognized. The apparent height of all features is depicted in a histogram in Figure 8b. A clear peak exists at 150 pm due to the features attributed to single atoms, with the shoulder at larger apparent heights originating from clusters. Considering that each cluster contains several Ir atoms, the coverage of the ≈150 pm high features agrees nicely with the assignment to single atoms. If the smallest species were dimers, our QCM calibration would have to be significantly underestimating the amount of Ir deposited. Increasing the Ir coverage to 0.05 ML (Figure 8c) increases the density of clusters but does not affect the density of adatoms.

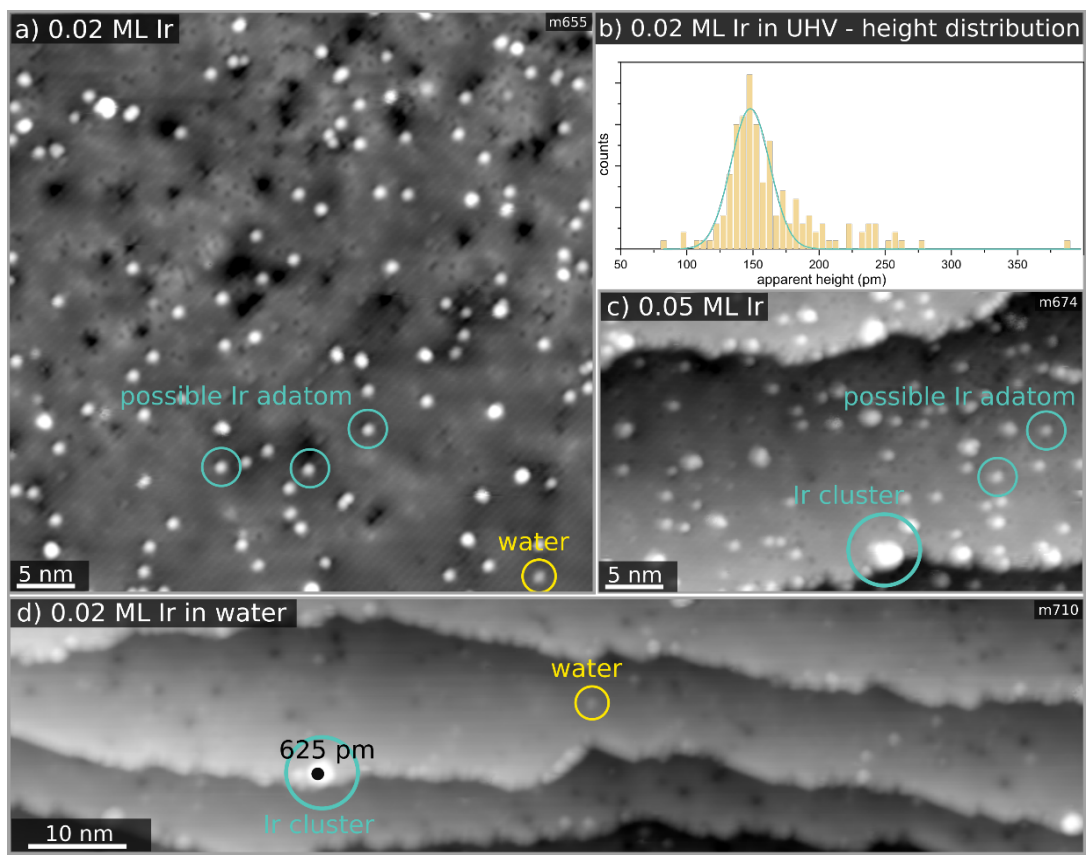

Figure 8: STM results of Ir on a-TiO$_2$(101). Frame (a) shows 0.02 ML Ir deposited in UHV ($V_{sample}$ = +1.5 V, $I_{tunnel}$ = 0.2 nA) and (b) the corresponding distribution of apparent heights. (c) 0.05 ML Ir deposited in UHV ($V_{sample}$ = +1.5 V, $I_{tunnel}$ = 0.2 nA) and (d) 0.02 ML Ir in water vapor background of 2×10$^{-8}$ mbar ($V_{sample}$ = +2 V, $I_{tunnel}$ = 0.2 nA). The highest cluster is marked in (d) and measures 625 pm at its highest point.

Figure 8d) shows the influence of water on the Ir/a-TiO$_2$(101) system. Deposition in water at room temperature leads to complete sintering of the single Ir atoms and the formation of large clusters at the step edges. This de-stabilizing effect of water is different to all the other metals studied here.

We also performed XPS measurements on the four metals deposited in UHV and in water vapor. Figure 9 shows an overview. For Pt, Rh and Ni, the peaks are shifted towards higher binding energy than those of the respective pure metals in the bulk. Water did not cause any drastic peak shifts, but intensity changes consistent with the propensity of dispersion/cluster formation observed in STM. The peak maxima are marked with a dotted line.

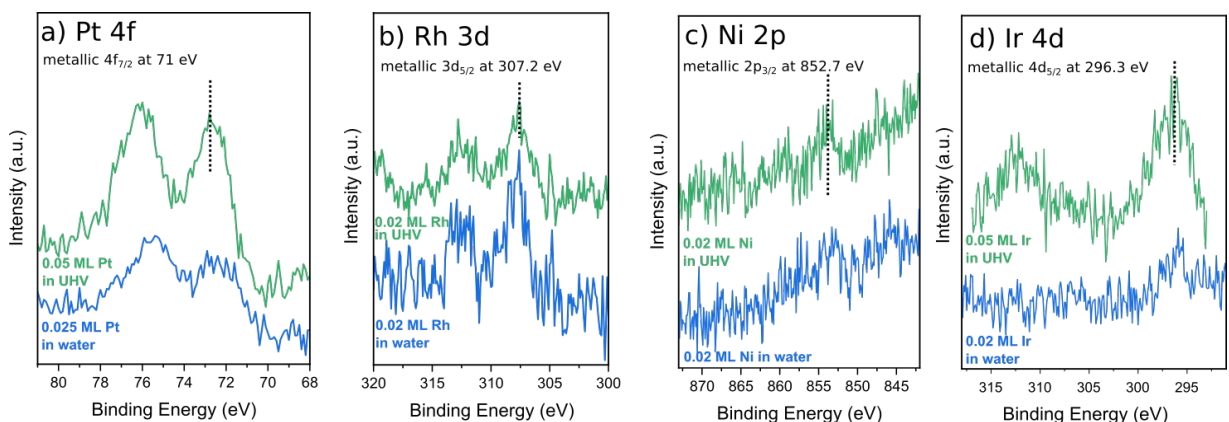

Figure 9: XPS spectra of Pt, Rh, Ni and Ir deposited in UHV and in a background of 2×10$^{-8}$ mbar water vapor.

## 4. Discussion

Overall, this study shows that Pt, Rh and Ni readily sinter after deposition onto the a-TiO$_2$(101) surface in UHV conditions. Rh is particularly unstable, and forms small clusters even at the lowest coverage studied with no evidence of any adatoms. Pt and Ni exhibit a mixture of small clusters and small, uniform features, which we assign as single atoms. Ir, in contrast, is highly dispersed at low metal coverages, but clusters begin to form when the coverage is increased. Our analysis of the adsorption site suggests that the adatom-assigned protrusions are between two surface O$_{2c}$ atoms for Pt and Ir, which is consistent with the site predicted for Pt by several DFT studies [34, 35, 38]. If the metal atoms bind to O, it is clear that the behaviour of the different metals can be understood in terms of the different oxygen affinities. Campbell and co-workers [39] recently studied adsorption of several late transition metals on MgO(110) and CeO$_{2-x}$(111), and reported the trend Ir > Ni > Pt > Rh for the oxygen affinity, which matches the relative stability observed for the UHV experiments here.

One issue with the assignment of adsorption at a regular lattice site is that the diffusion barrier for Pt atoms along the [010] direction has been calculated to be 0.86 eV [38]. Such a value means that Pt atoms could diffuse at room temperature on the ideal surface, which is likely why the majority of Pt atoms form small clusters before our STM measurements are conducted. Consequently, we infer that the immobile adatoms we observe at room temperature must be trapped at defect sites.

TiO$_2$ is sometimes considered synonymous with oxygen vacancies, because the behaviour of rutile TiO$_2$(110) [37] in UHV is dominated by $V_O$ sites. DFT calculations suggest that Pt atoms would indeed be highly stable at surface $V_O$ sites (4.71 eV, compared to 2.20 eV on the pristine surface) [34], but it is known that $V_O$s are preferentially accommodated in the subsurface layers. It is possible that such a large energy difference could cause a $V_o$ to diffuse to the surface[40] in the presence of Pt atoms, but this is inconsistent with our STM results for Pt and Ni:  In this case, the adatom would sit on an O$_{2c}$ site, not in-between as is consistently observed with Pt and Ir (Fig. 5). Consequently, we conclude that the immobile Pt and Ir species are stabilized by another defect type. Ni on the other hand is slightly shifted from the Pt and Ir adatoms and could therefore possibly be stabilized by a $V_o$.

The dark defects observed in STM are a primary candidate for the stabilization of metal adatoms because the defect is also located between two O$_{2c}$ atoms (compare Figures 5a and 5f). However, since the density of these defects is very inhomogeneous, which hinders any analysis of the number of defects covered by other species, we cannot exclude that another defect also plays a role. In any case, the nature of the dark defect is not clear. The previous assignment to substitutional Nb dopants [29] seems unlikely given the diffusion behaviour observed in the presence of water (Figure 2). Similar logic leads us to exclude that the defect is linked to substitutional Fe cations, although we do observe a small Fe2p signal in XPS survey spectra as this metal is a common contaminant in natural crystals. Nevertheless, Fe tends to be localized in patches on the surface, and its appearance differs significantly from the dark defects at standard imaging conditions.[26] Finally, we can also exclude that the dark defects are mistaken for an adsorbate, because the appearance of most candidate molecules present in the residual gas (water, $O_2$, CO, OH) has already been established on the a-$TiO_2$(101) surface. [29, 31] We propose that the defect is most likely a dopant atom present in an interstitial site in the lattice. Hebenstreit et al. speculated that it could possibly be a Ti interstitial.[41] While we cannot positively identify the chemical nature of the dark defect at this stage, our results suggest that extrinsic doping of the oxide could be a strategy to provide stronger binding sites capable of immobilizing expensive metals on a-$TiO_2$(101).

Turning now to the effect of water, we first note the completely different behaviour of Rh on a-$TiO_2$(101) compared to our previous study on α-$Fe_2O_3$(1$\bar{1}$02). In that work, depositing the metal in a background of water led to complete dispersion because Rh adatoms were stabilized by additional OH ligands [12]. One possible difference here is that water is already partially dissociated on α-$Fe_2O_3$(1$\bar{1}$02) at room temperature [42], so OH ligands are more readily available than on a-$TiO_2$(101) where water adsorbs molecularly. Another difference is the surface geometry: On α-$Fe_2O_3$(1$\bar{1}$02), OH groups adsorbed on nearby surface Fe cations can create a square planar environment for the $Rh_1$ adatom [12], which is known to be energetically favourable.[43] On a-$TiO_2$(101), this is not possible because OH groups adsorbed on surface Ti cations can at best create a threefold coordination, assuming the $Rh_1$ remains coordinated to two surface O atoms.

The only case where the dispersion seems to be aided by water is Ni, and our data suggest that ≈40 % of the deposited metal can be stabilized as isolated adatoms. The apparent height and adsorption site is the same as it was in the absence of water, so it is possible that some of the $Ni_1$ were already partially stabilized by the water inadvertently present in the residual gas in the UHV experiments. The complete opposite effect was seen after the deposition of Ir in a water vapor background. The presence of water promotes a dramatic sintering of the dispersed species, leading to mobile clusters, which finally get trapped at the steps. Clearly, then, the effect of water is difficult to predict, and given that water and hydroxyl groups are always present on metal-oxide surfaces, its omission from computational modelling of SAC systems is likely a major oversimplification. It is also important to recognise that water can have a significant effect on the reactivity, and there is evidence that water can play a role in SAC reaction mechanisms [44, 45].

Finally, one of the goals of this study was to assess the suitability of a-$TiO_2$(101) surface as a model support for surface science studies of SAC mechanisms. While it would be possible to study adsorption at the adatoms by STM/nc-AFM, the ambiguity over the nature of the defect sites that stabilize the adatoms precludes reliable modelling of the system. In any case, the presence of clusters at low coverage will make it difficult to distinguish the reactivity of single atoms from clusters using area averaging-techniques. At present, it is difficult to recommend this system as a suitable model system for studies of single-atom catalysis.

## 5. Conclusions

We have carried out room-temperature STM measurements of the a-$TiO_2$(101) surface after deposition of Pt, Rh, Ni and Ir in UHV and in a water vapor background. Pt and Ni form a mixture of small clusters and, possibly, single atoms. Rh exclusively forms clusters, while Ir is highly dispersed at a low coverage. The influence of water strongly varies from metal to metal. No influence is discernible for Pt and Rh, but the dispersion of Ni is increased when deposition is performed in a water vapor background. The exact opposite effect occurs in the case of Ir, which rapidly sinters after deposition in a water vapor background. The adsorption site of the species attributed to Pt and Ir atoms is the same as calculated for $Pt_1$ on the pristine surface; nevertheless, there is evidence that the single metal atoms are trapped at defect sites on the a-$TiO_2$(101) surface. As such, we conclude that doping of oxide surfaces could be a viable strategy to provide strong adsorption sites for single metal atoms.

Acknowledgement: LH, KD, and GSP acknowledge funding from the European Research Council (ERC) under the European Union's Horizon 2020 research and innovation programme (grant agreement No. [864628], Consolidator Research Grant "E-SAC"). UD acknowledges funding from the European Research Council (ERC) under the European Union's Horizon 2020 research and innovation programme grant agreement No. [883395], Advanced Research Grant 'WatFun'.